\documentclass{ieeeaccess}
\usepackage{cite}
\usepackage{amsmath,amssymb,amsfonts}
\usepackage{algorithmic}
\usepackage{graphicx}
\usepackage{textcomp}
\usepackage{xcolor}
\usepackage{soul}
\usepackage{wrapfig}

\usepackage{bm}
\makeatletter
\AtBeginDocument{\DeclareMathVersion{bold}
\SetSymbolFont{operators}{bold}{T1}{times}{b}{n}
\SetSymbolFont{NewLetters}{bold}{T1}{times}{b}{it}
\SetMathAlphabet{\mathrm}{bold}{T1}{times}{b}{n}
\SetMathAlphabet{\mathit}{bold}{T1}{times}{b}{it}
\SetMathAlphabet{\mathbf}{bold}{T1}{times}{b}{n}
\SetMathAlphabet{\mathtt}{bold}{OT1}{pcr}{b}{n}
\SetSymbolFont{symbols}{bold}{OMS}{cmsy}{b}{n}
\renewcommand\boldmath{\@nomath\boldmath\mathversion{bold}}}
\makeatother

\def\BibTeX{{\rm B\kern-.05em{\sc i\kern-.025em b}\kern-.08em
    T\kern-.1667em\lower.7ex\hbox{E}\kern-.125emX}}

\begin{document}
\makeatletter
\def\@doi{}
\makeatother


\title{Learning to Diagnose Privately: DP-Powered LLMs for Radiology Report Classification
}
\author{Payel Bhattacharjee \authorrefmark{1},
Fengwei Tian \authorrefmark{1}, Geoffrey D. Rubin \authorrefmark{2}, Joseph Y. Lo \authorrefmark{3}, Nirav Merchant, \authorrefmark{4}, Heidi Hanson \authorrefmark{5}, John Gounley \authorrefmark{5}, Ravi Tandon \authorrefmark{1} }

\address[1]{Department of Electrical and Computer Engineering, University of Arizona, AZ, USA (e-mail: \{payelb,fengtian,tandonr\}@arizona.edu)}
\address[2]{Department of Radiology $\&$ Imaging Sciences, University of Arizona, Tucson, AZ, USA (email: grubin@arizona.edu)
}
\address[3]{Departments of Radiology and Electrical and Computer Engineering, Duke University, Durham, NC, USA (email: joseph.lo@duke.edu).}
\address[4]{Data Science Institute, University of Arizona, Tucson, AZ, USA (email: nirav@arizona.edu)}
\address[5]{Oak Ridge National Laboratory, Oak Ridge, TN, USA (email: \{hansonha, gounleyjp\}@ornl.gov ) }

\tfootnote{This work was supported by NIH Award R01-CA261457-01A1 and also by the US Department of Energy, Office of Science, Office of Advanced Scientific Computing under Award Number DE-SC-ERKJ422. US NSF under Grants CCF-2100013, CNS2209951, CNS-2317192.
}

\corresp{Corresponding author: Payel Bhattacharjee (e-mail: payelb@arizona.edu).}

\begin{abstract} Large Language Models (LLMs) are being widely adopted in different domains including education, healthcare, and finance. In healthcare domain, LLMs are used in disease diagnosis, abnormality classification, remedy suggestions etc.. Multi-abnormality classification of radiology reports is essential in healthcare, medical decision-making, and drug discovery. LLMs are increasingly utilized for such tasks because of their remarkable Natural Language Processing (NLP) capabilities, which streamline medical report processing and reduce administrative burden. To enhance the predictive accuracy, LLMs are often fine-tuned on private, locally available datasets, such as medical reports. However, this practice raises significant privacy concerns, as LLMs are prone to memorizing training data, making them susceptible to data extraction attacks even through query-based access. Additionally, sharing fine-tuned models and weights poses adversarial risks, because they may inadvertently reveal sensitive information about the training data. Despite the growing application of LLMs to medical text classification, privacy-preserving fine-tuning for multi-abnormality classification remains underexplored. To bridge this gap, we propose a differentially private (DP) fine-tuning approach that preserves privacy while enabling multi-abnormality classification from text radiology reports through Low Rank Adaptation (LoRA). Our framework leverages DP optimization techniques to fine-tune LLMs on local patient data while mitigating data leakage risks. To our knowledge, this is the first study to incorporate DP fine-tuning of LLMs for multi-abnormality classification using text-based radiology reports. We used labels generated by a larger LLM to fine-tune a smaller LLM, accelerating inference while maintaining privacy constraints. We conducted extensive experiments on the MIMIC-CXR, and CT-RATE datasets to evaluate DP fine-tuning method across varying privacy regimes, analyzing the privacy-utility trade-off and demonstrating the efficacy of our approach.  {For instance, on the MIMIC-CXR dataset, our proposed DP-LoRA framework achieves weighted F1-scores of up to 0.89 under moderate privacy budgets ($\epsilon = 10$), approaching the performance of non-private LoRA ($0.90$) and full fine-tuning ($0.96$). These results demonstrate that strong privacy protection can be achieved with only moderate performance degradation.}
\end{abstract}

\maketitle
\begin{figure*}[t]
    \centering
    \includegraphics[scale = 0.26]{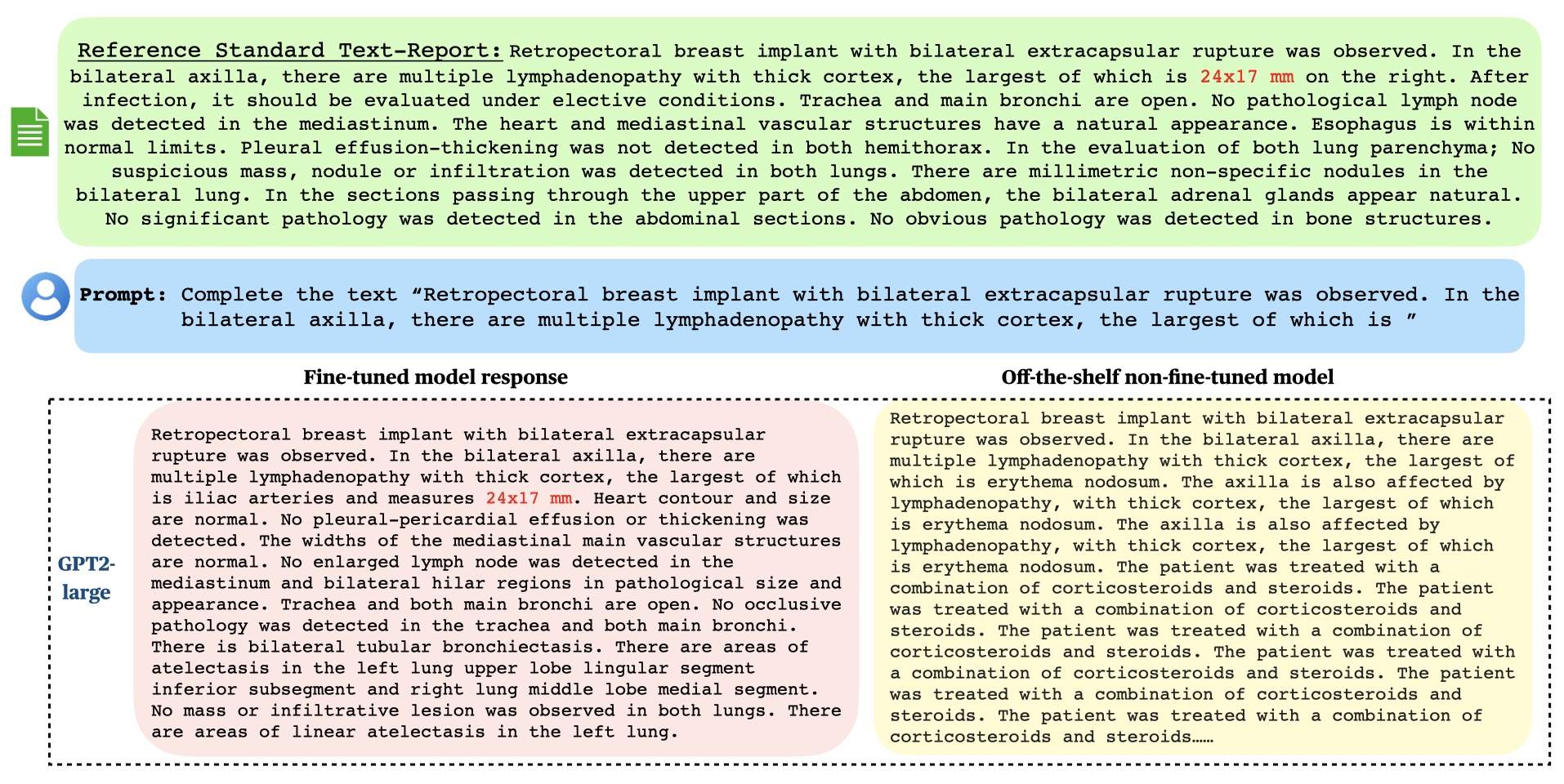}
    \caption{Memorization behavior of the GPT‑2 Large model on text‑based radiology reports reveals that, during report completion tasks, the model can recall and reproduce patient‑specific numerical values (e.g., 24 × 17 mm) from its training data, thereby amplifying privacy risks. }
    \label{fig:memorization_1}
\end{figure*}

\section{Introduction}
Innovations and advancements in Large Language Models (LLMs) continue to redefine the frontiers of artificial intelligence (AI), particularly in critical domains such as healthcare, and clinical diagnosis where precision, interpretability, and efficiency are paramount. Fine-tuning LLMs on task-specific, locally sourced, and often private datasets is a common approach for adapting models for specialized downstream applications. Although such datasets are typically de-identified to remove personally identifiable and HIPAA-protected health information (PHI), certain sensitive elements, such as device identifiers, medication names, or contextual patterns, may still be inadvertently memorized by models. When combined with external data sources, this memorization can amplify privacy risks \cite{shokri2017membership} by enabling partial data reconstruction or inference of sensitive attributes.

Recent research has demonstrated that AI models are prone to memorizing \cite{carlini_membership_2022,carlini_extracting_nodate} fragments of their training data,  which lead to significant privacy vulnerabilities. For instance, querying fine-tuned models can reveal personal details or verbatim text segments from the training corpus. During fine-tuning on clinical data such as free-text radiology reports, models may inadvertently encode and reproduce identifiable content. When cross-referenced with publicly available datasets, this leakage can facilitate attacks such as membership inference (MIA) or jailbreak prompting \cite{carlini_membership_2022,carlini_extracting_nodate,duan_membership_2024}. As highlighted in previous studies \cite{yan2024dp, akinci_dantonoli_cybersecurity_2025}, we underscore the inherent risks associated with fine-tuning LLMs on sensitive medical text and emphasize the necessity of \textit{Differentially Private Fine-Tuning} (DP-FT).  { To illustrate the practical privacy risks associated with fine-tuning on clinical data, we demonstrate a memorization example using a generative language model trained on radiology reports in the next sub-section.}

\begin{figure*}[t]
    \centering
    \includegraphics[scale = 0.25]{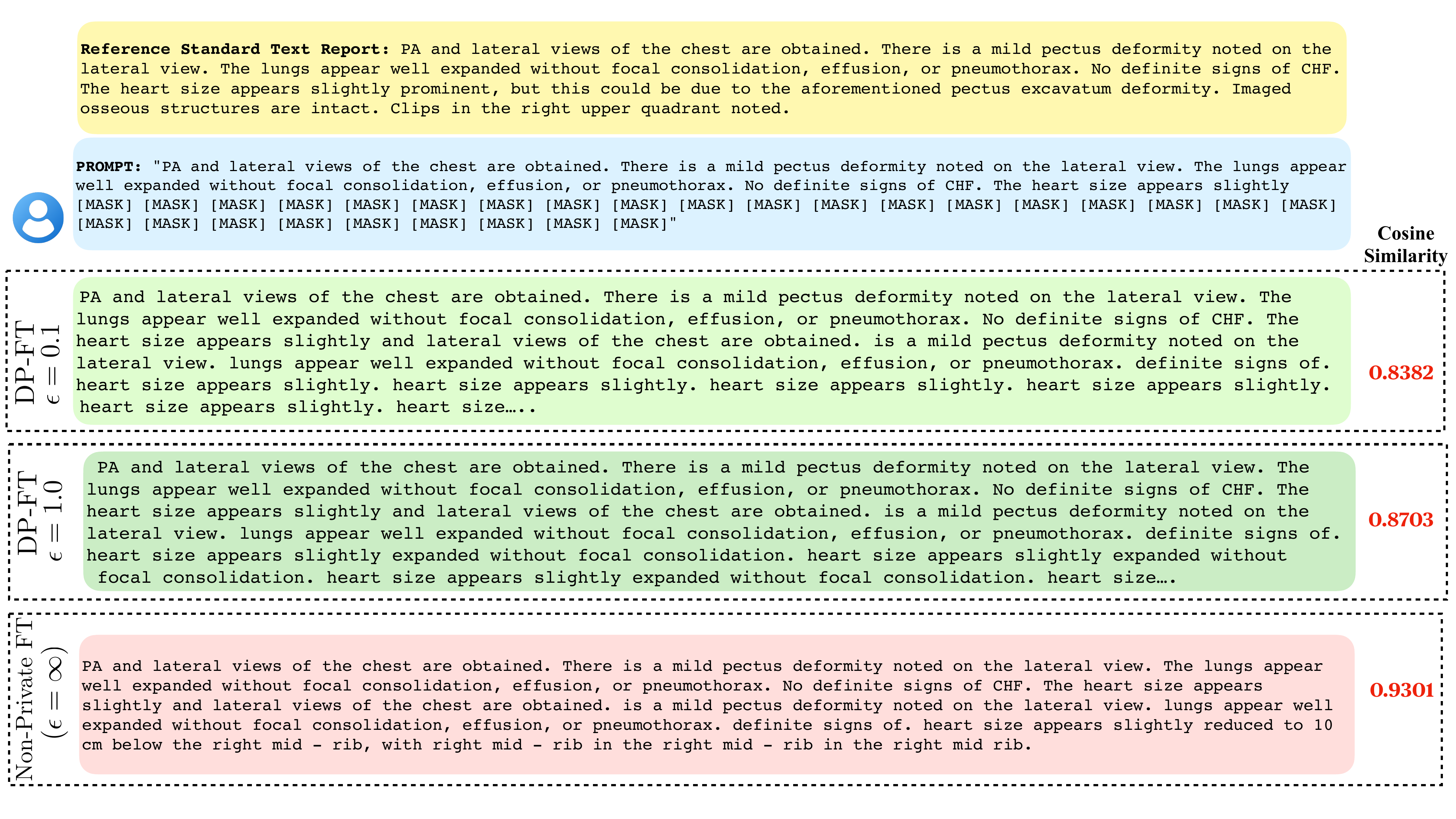}
    \caption{Example illustrates rising Cosine similarity between original report Findings and BERT-base-generated Findings, indicating memorization even in non-causal models, more pronounced in the fine-tuned variant.  { While the non-privately fine-tuned model gives highest cosine similarity (indicating most memorization), fine-tuning with DP reduces the cosine similarity score indicating lesser memorization and better privacy.}}
    \label{fig:memorization_2}
\end{figure*}

\textit{\textbf{Text-memorization illustration by LLMs:}} As mentioned previously (also shown in Figure \ref{fig:memorization_1}, and Figure \ref{fig:memorization_2}), fine-tuning LLMs can pose significant privacy risks when models are exposed to real patient data. In Figures \ref{fig:memorization_1}, \ref{fig:memorization_2}, we illustrate these threats via text-memorization using  BERT (used for abnormality classification), and GPT2 (causal language model), and show how these fine-tuned LLMs can memorize parts of training/fine-tuning data.  {The examples highlight that fine-tuned models can reproduce patient-specific measurements verbatim. This behavior demonstrates a direct privacy leakage mechanism and motivates the need for formally private fine-tuning methods rather than relying solely on dataset de-identification.} Our findings show fine-tuned LLMs reflect memorized training content compared to the off-the-shelf (non-fine-tuned) LLM, this indicates that fine-tuned models exhibit especially strong tendencies towards such leakage. As shown in Figure \ref{fig:memorization_2}, when asked to complete a partial report, fine-tuned GPT2 models were able to replicate parts of the original report “Findings” (24x17mm). Moreover, for some other examples, we observe that the fine-tuned model outputs the next patient’s report (for instance patient with patient identifier p1xxxxxxx) from the fine-tuning dataset, highlighting the privacy vulnerabilities of non-privately fine-tuned LLMs. In Figure \ref{fig:memorization_1}, we observe a similar trend for fine-tuned BERT model used for multi-abnormality classification. It is interesting to note that BERT based models are not causal language models, yet they still have the ability to replicate portions of reports used for fine-tuning. Specifically, the fine-tuned models achieved a higher cosine similarity score (between generated and original report “Findings”) privately fine-tuned ones (refer to Figure \ref{fig:avg_cosine}). Cosine similarity score measures how similar two pieces of text are by comparing the angle between their vector representations in the embedding space; a higher cosine similarity between generated text and the original training data suggests the model may be closely repeating what it saw during training, indicating memorization. This underscores the potential privacy risks posed by LLMs memorizing sensitive portions of medical reports, especially when applied to tasks such as text completion, disease classification, report summarization, and more.  

While prior studies have explored differential privacy (DP) in model training, its integration into LLM fine-tuning for text-based radiology report classification remains underexplored.  {To the best of our knowledge, this is among the first systematic empirical studies evaluating differentially private LoRA-based fine-tuning for multi-abnormality classification from free-text radiology reports}. Motivated by previous works \cite{hong_dp-opt_2024, liu_differentially_2024} who presented DP-based fine-tuning of LLMs for medical image classification, we propose a privacy-preserving framework that incorporates \textit{Low-Rank Adaptation} (LoRA) to enable secure fine-tuning of LLMs on domain-specific text-based radiology data. This framework is adaptable to any task involving free-text medical reports while ensuring formal differential privacy guarantees during the fine-tuning process. 

 {Despite the rapid adoption of LLMs in healthcare text processing, existing works largely address either predictive performance or privacy protection independently. Most medical natural language processing (NLP) related studies optimize classification accuracy without formal privacy guarantees, while privacy-focused research typically evaluates generic datasets rather than highly sensitive clinical narratives such as radiology reports. Furthermore, the interaction between differential privacy and parameter-efficient fine-tuning strategies remains insufficiently understood. Consequently, it is unclear whether practical privacy-preserving LLM adaptation can maintain clinically meaningful performance. This paper addresses this gap by systematically evaluating differentially private LoRA-based fine-tuning for multi-abnormality classification from free-text radiology reports.}

\section{Related Works}
Modern machine learning (ML), deep learning (DL), and natural language processing (NLP) techniques have greatly advanced multi-abnormality classification, particularly in medical imaging and radiology report interpretation. High-dimensional Computed Tomography (CT) and volumetric imaging data \cite{hamamci_developing_2025} are routinely analyzed for abnormality detection, while traditional deep learning frameworks have achieved substantial success in multi-lesion recognition, abdominal disease detection, and tumor identification. Weakly supervised methods \cite{tushar2021classification} further enhance classification accuracy by reducing reliance on extensive labeled datasets.

Beyond imaging, radiology report generation and text-based classification have emerged as critical tasks, leveraging a combination of ML, DL, and rule-based algorithms. The advent of LLMs has transformed medical text processing, particularly for summarization and classification. Extractive summarization methods using models such as BERTSUM \cite{liu_fine-tune_2019}, TransEXT, and BERT \cite{liu_fine-tune_2019} have proven effective for medical narrative analysis. Domain-specific LLMs including BioBERT \cite{lee_biobert_2020}, RadioBERT \cite{kaur_radiobert_2022}, and ChestXRayBERT \cite{cai_chestxraybert_2023} further enhance radiology report understanding, while models such as ChexBERT, Chexpert, and Vicuna facilitate multi-abnormality classification through contextualized embeddings.

Despite their promise, fine-tuning large-scale LLMs for medical applications remains computationally intensive due to the vast number of model parameters. Parameter-efficient tuning methods \cite{hu_lora_2021} such as LoRA, Adapters, and Compacters have been developed to mitigate computational overhead while maintaining performance. However, fine-tuning on private clinical datasets introduces privacy concerns, as sensitive patient information can be inadvertently exposed. To address these risks, prior studies have proposed differentially private frameworks for federated and multi-modal learning \cite{hsu_differential_2014, hong_dp-opt_2024}. Recent work on differentially private optimization (DP-OPT) \cite{hong_dp-opt_2024} and private prompt tuning underscores the importance of privacy-preserving approaches in medical AI. Recent advances, such as DP-SSLoRA \cite{yan2024dp}, leverage self-supervised representation learning and low-rank reparameterization to enhance the privacy–utility trade-off in differentially private medical models. This work demonstrates that integrating self-supervision with DP-LoRA can achieve strong privacy guarantees without sacrificing diagnostic accuracy.

In this paper, we focus on multi-abnormality classification using free-text radiology reports. While LLMs excel at text processing, their lack of specialized medical training and the high computational cost of fine-tuning limit their performance in clinical tasks. By integrating differential privacy with parameter-efficient tuning via LoRA, we aim to reduce fine-tuning complexity while maintaining both privacy and predictive accuracy.  {In summary, prior literature typically studies either performance optimization of medical NLP models or privacy vulnerabilities in isolation. Comprehensive evaluations that jointly analyze performance and privacy in sensitive clinical text remain limited. Radiology reports represent highly sensitive medical narratives where memorization risks are critical but underexplored. Moreover, differential privacy mechanisms combined with parameter-efficient fine-tuning frameworks such as LoRA have not been systematically evaluated for multi-abnormality classification tasks. This work bridges these gaps by providing a unified empirical and methodological study of privacy-preserving LLM adaptation in clinical text analysis.}

\begin{figure*}[t]
    \centering
    \includegraphics[scale = 0.45]{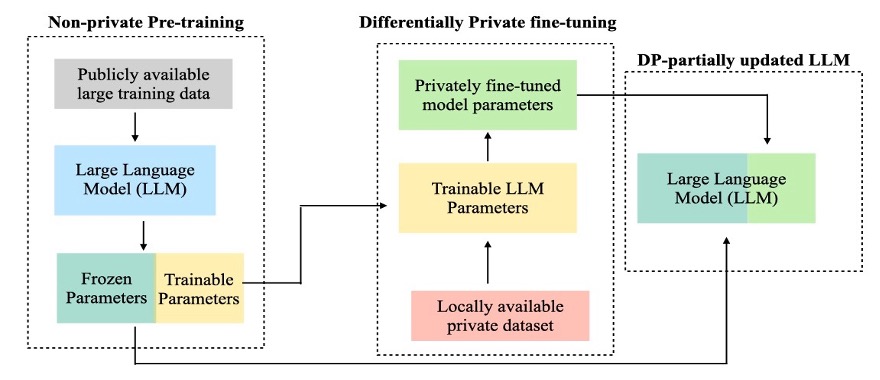}
    \caption{Differentially private fine-tuning updates only a subset of low-rank model parameters on local medical data, while keeping the rest of the pre-trained LLM frozen.}
    \label{generic_workflow}
\end{figure*}
\section{Objective and Contributions}
Leveraging multi-abnormality labels generated from free-text radiology reports for differentially private fine-tuning (DP-FT) of LLMs provides a promising pathway for achieving efficient and privacy-preserving inference. However, this strategy remains underexplored for multi-abnormality classification from free-text radiology reports. To the best of our knowledge, no prior work has incorporated DP-LoRA~\cite{liu_differentially_2024} for multi-abnormality classification from text radiology reports.

In this study, we first evaluate LoRA-based fine-tuning under non-private conditions to establish baseline performance. While non-private fine-tuning yields strong accuracy, it also increases the risk of memorizing sensitive data. To mitigate this, we propose \textit{Differentially Private Low-Rank Adaptation} (DP-LoRA), a fine-tuning approach that integrates LoRA with the privacy guarantees of differential privacy (DP). Specifically, we employ \textit{Differentially Private Stochastic Gradient Descent} (DP-SGD)~\cite{Abadi2016}, where gradients are clipped to a threshold $C$ and perturbed with Gaussian noise characterized by privacy parameters $(\epsilon, \delta)$. This ensures formal and quantifiable protection against data leakage while maintaining competitive classification performance.

\section{Preliminaries}
In this Section, we provide some preliminaries on the datasets, and models we have used in this paper along with some background on differential privacy and low rank adaptation.
\subsection{Datasets}
In this study, we have utilized two publicly available, widely used radiology datasets: the \textit{MIMIC-CXR} \cite{johnson_mimic-cxr_2019}, and the \textit{CT-RATE} dataset \cite{hamamci_developing_2025}. MIMIC-CXR dataset comprises 227,835 radiographic studies collected from the Beth Israel Deaconess Medical Center in Boston, MA, USA.  {Sub-sampling improves privacy guarantees only under randomized sampling assumptions. Our implementation follows standard DP-SGD mini-batch sampling, where each sample is independently included with a fixed probability. In this setting, privacy amplification by sub-sampling reduces the effective privacy loss per iteration.} Therefore, we use a subset of the MIMIC-CXR dataset, specifically 20,883 text reports for training and 2,610 reports for testing. { We selected a fixed subset of 20,883 training reports to ensure consistent privacy accounting under DP-SGD. Differential privacy guarantees depend on dataset size and sampling rate; therefore, controlling dataset cardinality allows reproducible and interpretable privacy budgets across experiments.} For patients with multiple radiographic studies, we aggregate all associated textual reports, creating unified “Findings” and “Impression” sections to ensure a comprehensive representation.  {The “Findings” and “Impression” sections are aggregated because clinical diagnoses may appear in either section. Combining them ensures that all abnormality evidence is captured, improving label consistency for multi-label classification.} Additionally, duplicate entries, if any, are removed. MIMIC-CXR has been widely used as a benchmark dataset in prior research. However, to the best of our knowledge, no existing work has incorporated differentially private fine-tuning of LLMs for multi-abnormality classification. While we apply this framework to MIMIC-CXR, the proposed differentially private fine-tuning approach can be adapted to other medical text report datasets for multi-abnormality classification.
The CT-RATE dataset comprises 25,692 non-contrast chest CT scans, sourced from 21,304 unique patients, and the CT scans are expanded to 50,188 volumes through various reconstructions. Each scan is accompanied by associated radiology reports, multi-abnormality labels, and metadata. The patient cohort was partitioned into two subsets: 20,000 patients were assigned to the training set, while 1,304 patients were designated for validation. We have used the entire CT-RATE dataset, and utilized 80$\%$ for training, 10$\%$ for validation and 10$\%$ for testing.  {Although CT-RATE contains reconstruction-expanded volumes, we partition the dataset at the patient level, preventing leakage of correlated samples across training and testing splits and avoiding evaluation bias.}

\subsection{Models and Reference Standards}
\begin{figure*}[t]
    \centering
    \includegraphics[scale = 0.52]{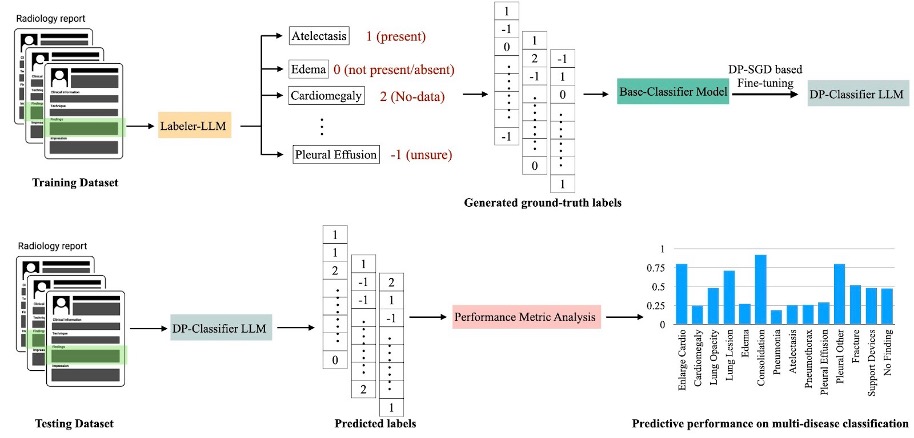}
    \caption{Workflow of the proposed DP fine-tuning framework for multi-abnormality classification problem from the “Findings” of free-text chest radiology reports.}
    \label{fig:framework}
\end{figure*}

Figures \ref{generic_workflow}, and \ref{fig:framework} illustrate the workflow of the proposed private multi-abnormality classification framework. 
For MIMIC-CXR, ChexBERT \cite{smit_chexbert_2020} was used to generate 14 diagnostic labels: 
\textit{Enlarged Cardiomediastinum, Cardiomegaly, Lung Opacity, Lung Lesion, Edema, Consolidation, Pneumonia, Atelectasis, Pneumothorax, Pleural Effusion, Pleural Other, Fracture, Support Devices, and No Finding}. 

For CT-RATE, the RadBERT \cite{yan_radbert_2022} model performed binary classification (present/absent) over 18 classes, including 
\textit{Cardiomegaly, Pericardial Effusion, Coronary Artery Wall Calcification, Lymphadenopathy, Lung Nodule, Pulmonary Fibrotic Sequela, and Pleural Effusion}.  {The selected models (ALBERT-base, BERT-small, and BERT-medium) were chosen to represent compact transformer architectures with varying parameter sizes (11M-41.7M parameters). In differentially private optimization, the magnitude of injected noise depends on model dimensionality. Therefore, smaller models offer improved privacy-utility trade-offs and are more suitable for clinical deployment environments with limited computational and memory resources. Our study evaluates this relationship by comparing multiple model capacities under identical privacy budgets.}

\subsection{Differential Privacy}
Differential privacy (DP) provides a statistical guarantee that the inclusion or exclusion of any single individual's data does not significantly affect the model output. 

Formally, \textit{a randomized algorithm $\mathcal{M}$ with input domain $\mathcal{X}$ and output range $\mathcal{R}$ is $(\varepsilon,\delta)$-differentially private if, for all neighboring datasets $D$ and $D'$ differing by one record, and for all subsets $S \subseteq \mathcal{R}$,}
\begin{equation}
\mathbb{P}[\mathcal{M}(\mathcal{D}) \in S] \le e^{\epsilon} \cdot \mathbb{P}[\mathcal{M}(\mathcal{D}') \in S] + \delta.
\end{equation}
Smaller $\epsilon$ implies stronger privacy, while $\delta$ allows for a negligible probability of privacy loss.

\begin{figure*}
    \centering
    \includegraphics[scale = 0.25]{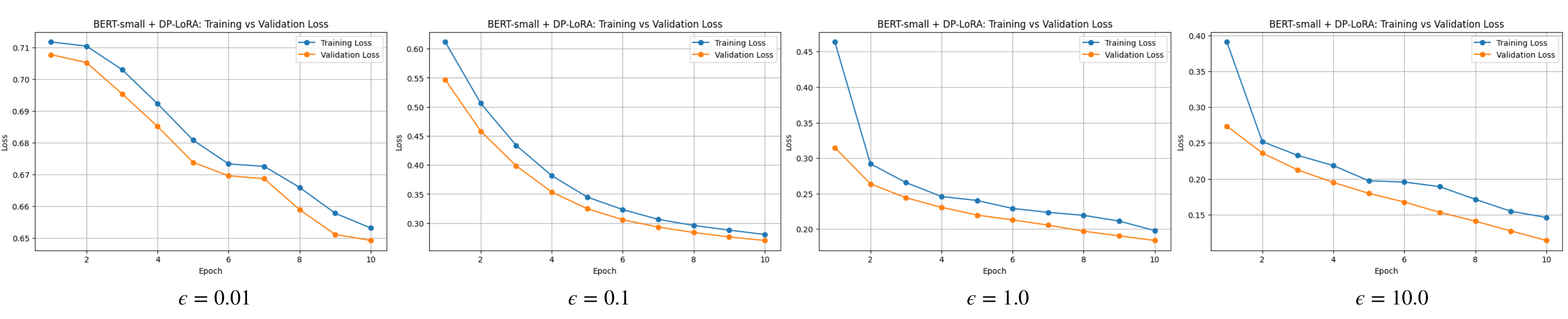}
    \caption{ {Training and Validation loss of BERT-small model with varying $\epsilon$ values on MIMIC-CXR dataset. The plots highlight there is no evidence of overfitting.}}
    \label{fig:bertsmall_loss}
\end{figure*}

\subsection{Low-Rank Adaptation (LoRA)}
Low-Rank Adaptation (LoRA) \cite{hu_lora_2021} is a parameter-efficient fine-tuning method that avoids updating the full weight matrix $W \in \mathbb{R}^{d \times k}$. 
Instead, it introduces a low-rank decomposition:
\begin{equation}
W' = W + BA,
\end{equation}
where $B \in \mathbb{R}^{d \times r}$, $A \in \mathbb{R}^{r \times k}$, and $r \ll \min(d, k)$. 
Only low-rank matrices $A$ and $B$ are updated during fine-tuning, while $W$ remains frozen. 
This reduces the number of trainable parameters and hence the overall computational overhead significantly reduces.

\subsection{LoRA with Differential Privacy}
To ensure differential privacy during fine-tuning, we employ \textit{Differentially Private Stochastic Gradient Descent} (DP-SGD) \cite{Abadi2016} in the pipe-line of LoRA. 
In DP-SGD, at each optimization step, the gradient $\nabla \theta$ is clipped to a fixed norm bound $C$ to limit sensitivity:
\begin{equation}
\tilde{g} = \frac{g}{\max(1, \frac{\|g\|_2}{C})}.
\end{equation}
Gaussian noise with standard deviation $\sigma = 1.25 \times (1/\epsilon)$ is then added:
\begin{equation}
g' = \tilde{g} + \mathcal{N}(0, \sigma^2 C^2 \mathbf{I}),
\end{equation}
ensuring $(\epsilon, \delta)$-differential privacy over the entire fine-tuning process. 
By updating only low-rank parameters, DP-LoRA minimizes the number of noisy updates while maintaining formal privacy guarantees and strong predictive performance.

\subsection{Selection of Privacy Parameters}
The selection of $(\epsilon, \delta)$ governs the privacy–utility trade-off. 
Smaller values yield stronger privacy but may reduce model accuracy. 
In healthcare applications, $\varepsilon \in (0, 10)$ is typical for robust protection \cite{hsu2014differential}. 
An appropriate choice for $\delta$ is $1/n^2$, where $n$ denotes the number of training samples, ensuring that the probability of privacy failure decreases with dataset size.  {We evaluate $\epsilon \in  \{0.01, 0.1, 1, 10 \}$ to represent four privacy regimes: very strong, strong, moderate, and weak privacy respectively. These ranges follow commonly adopted values in medical differential privacy studies. Rather than exhaustive sweeping, this selection provides interpretable operating points for practitioners to balance privacy and performance.}
\begin{figure*}[t]
    \centering
    \includegraphics[scale = 0.25]{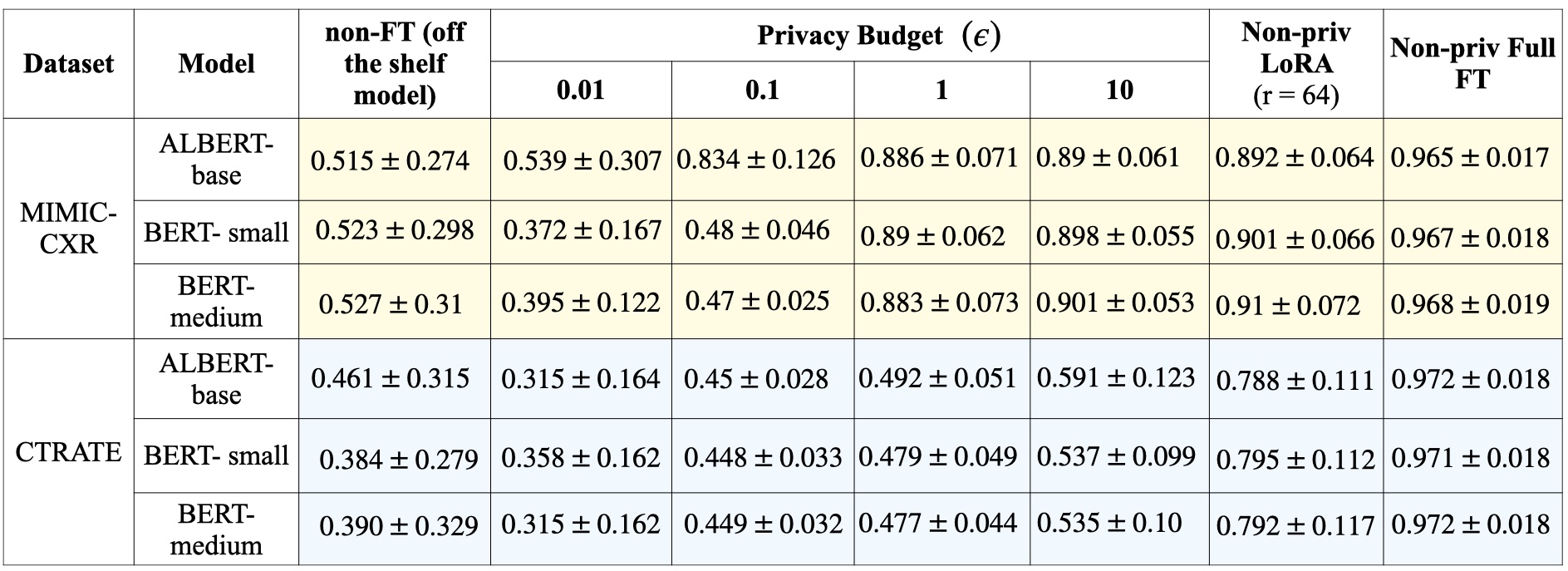}
    \caption{This table shows the mean and standard deviation of the weighted F1-score (over 10 epochs) of 3 models BERT-small, BERT-medium, and ALBERT-base over different privacy regimes with four different privacy budgets = (0.01, 0.1, 1.0, 10.0).  This shows the privacy-utility trade-off in different privacy regimes.
}
    \label{fig:results_table}
\end{figure*}
\begin{figure*}[t]
    \centering
    \includegraphics[scale = 0.36]{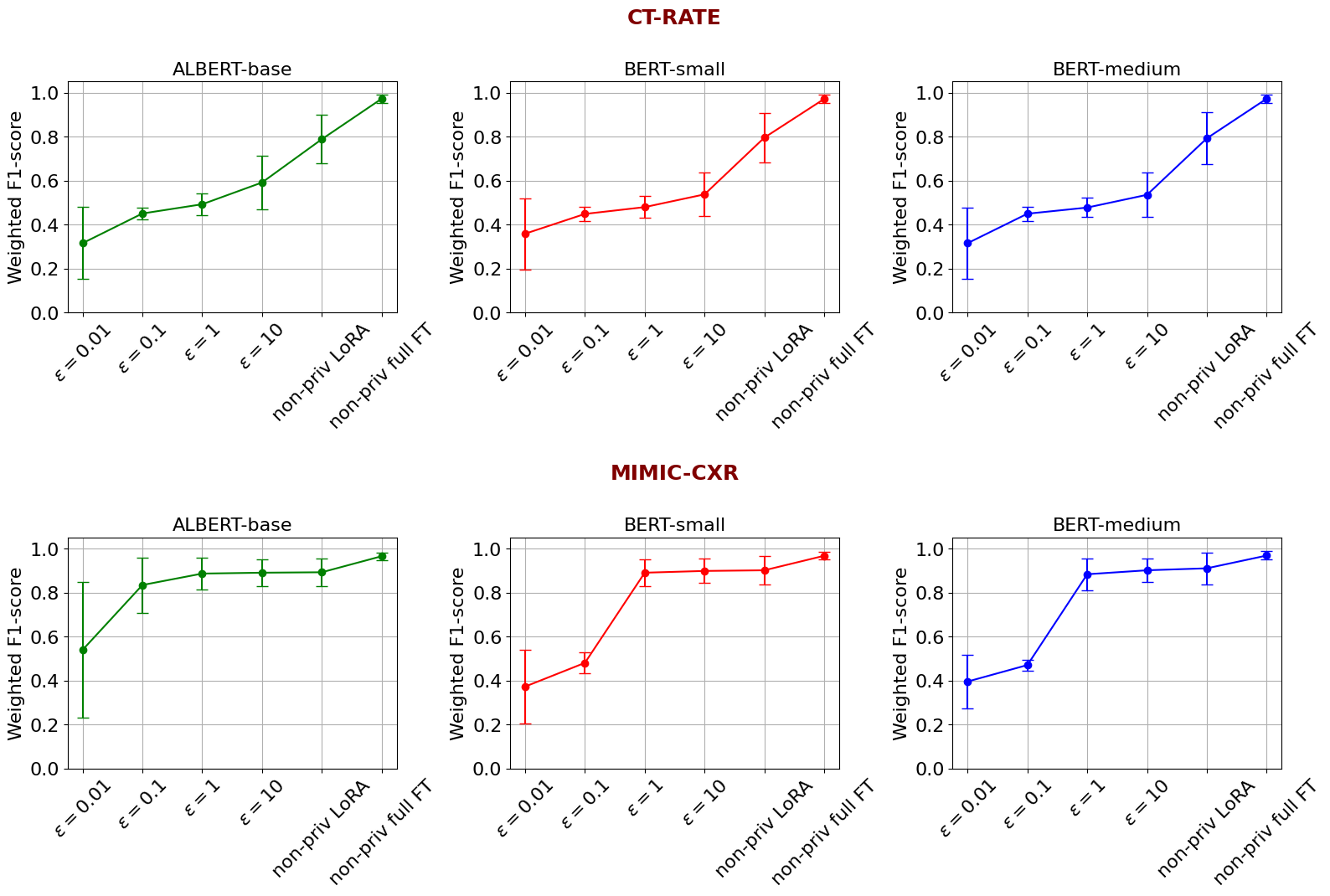}
    \caption{Privacy-utility tradeoff for ALBERT-base, BERT-small and BERT-medium models across varied privacy regimes. As the privacy budgets increase, due to less stringent privacy constraints, the model utility increases.
}
    \label{fig:f1score}
\end{figure*}

\section{Technical Contributions}

In this study, we propose a privacy-preserving framework for multi-abnormality classification from free-text radiology reports by integrating LLM-based weak supervision, Low-Rank Adaptation (LoRA), and Differentially Private Stochastic Gradient Descent (DP-SGD). Figure 3 highlights the broader overview of DP-LoRA framework. Figure \ref{fig:framework} illustrates the overall workflow, which consists of three main stages: \textit{(1) label generation using a domain-specific labeler LLM, (2) differentially private fine-tuning of a classifier LLM, and (3) performance evaluation of a held-out test dataset.}

\subsection{Weakly Supervised Label Generation}
We first utilize specialized labeler LLMs (e.g., CheXbert for MIMIC-CXR dataset and RadBERT for CT-RATE dataset) to automatically extract structured abnormality labels from free-text radiology reports. Each report is mapped to a multi-label vector across clinically relevant findings such as atelectasis, edema, cardiomegaly, and pleural effusion and so on.  
The raw outputs of ChexBERT are standardized into a unified label representation, where: $(+1)$ indicates disease presence, $(0)$ indicates disease absence or no relevant mention,  $(-1)$ represents uncertain findings, $(+2)$ denotes missing or no-data entries.
This stage enables scalable, annotation-free label generation while maintaining full compatibility with privacy-preserving downstream training. The labels for MIMIC-CXR dataset were binarized by merging “unsure” ($-1$) and “no data” ($+2$) into the negative ($0$) class, and retaining “present” ($+1$) as positive. For CT-RATE dataset, the RadBERT generated labels have binary labels, present/absent $(0/+1)$.

\noindent {\textbf{Labeler Bias Consideration:} The labels are generated using CheXbert and RadBERT, which share domain similarity with the evaluation task. This may introduce inductive bias where the classifier partially learns the labeler’s representation rather than purely clinical semantics. However, three factors mitigate this effect: (1) the classifier models (ALBERT/BERT variants) differ architecturally from the labeling models, (2) evaluation uses held-out patient partitions, and (3) DP noise discourages exact pattern imitation. Therefore, the reported performance reflects robust feature learning rather than simple labeler mimicry, though future work with expert annotations would further strengthen validation.}

\subsection{Differentially Private Fine-Tuning of the Classifier LLM}
The generated multi-label vectors (as shown in Figure \ref{fig:framework}) serve as supervision signals for fine-tuning a base classifier LLM using differentially private optimization. Rather than updating the full parameter set, we adopt DP-LoRA, which inserts low-rank trainable matrices into the transformer layers and freezes all original model weights. In our proposed workflow since only the LoRA parameters are updated, the algorithm significantly reduces the number of noisy updates, improving model utility under strong privacy constraints.

\subsection{Inference and Performance Evaluation}
During inference, the DP-Classifier LLM predicts abnormality labels for unseen radiology reports using only the differentially private low-rank parameters.  
The predicted label vectors are compared with reference standards to compute disease-wise performance metrics, the weighted F1-score. As illustrated in Figure \ref{fig:f1score}, this enables fine-grained assessment across all abnormality classes and highlights categories most sensitive to privacy-induced noise.

\begin{figure*}[t]
    \centering
    \includegraphics[scale=0.26]{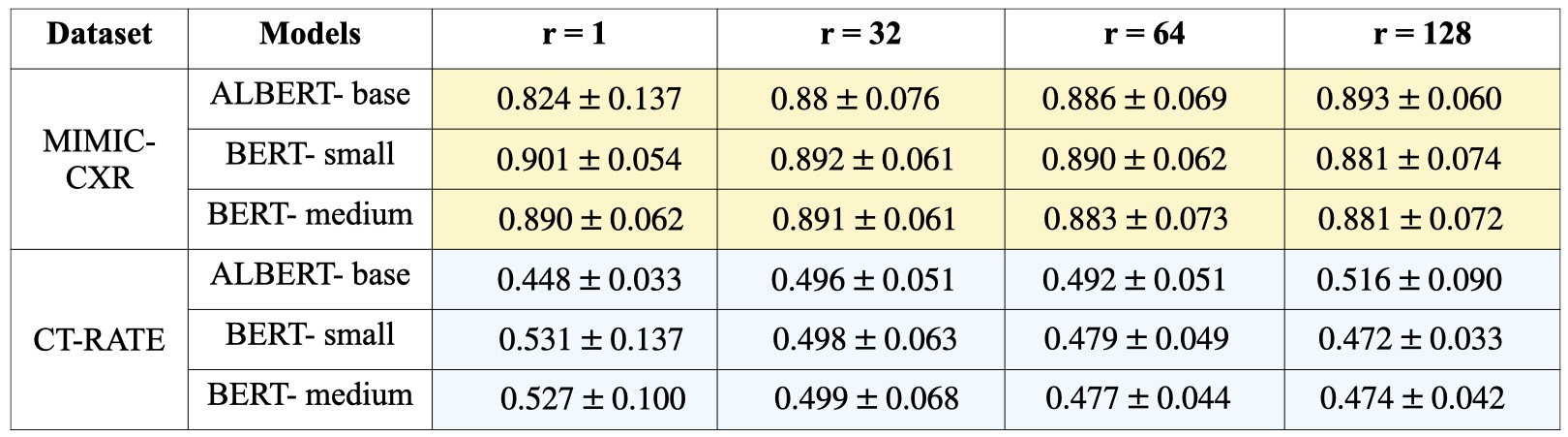}
    \caption{This table represents the predictive performance of 3 models ALBERT-base, BERT-small, and BERT-medium for a fixed privacy budget of = 1.0, and DP-failure probability of   for varied LoRA ranks (i.e., number of fine-tuned parameters).}
    \label{fig:lora}
\end{figure*}
\section{Results}
In this Section, we present the results on private and non-private fine-tuning of 3 different models on 2 different datasets. Specifically, we present the experimental results on the MIMIC-CXR \cite{johnson_mimic-cxr_2019} and CT-RATE \cite{hamamci_developing_2025} datasets with 3 models of different sizes: BERT-medium (41.7M parameters), BERT-small (29.1M parameters) \cite{bhargava2021generalization}, and ALBERT-base-v2 (11M parameters) \cite{lan_albert_2020}. For the scope of this paper we have considered the weighted F1-score \cite{erberk_uslu_nlp-powered_2024} metric while reporting the predictive performance of the fine-tuned models with and without differential privacy. To evaluate model performance in the presence of class imbalance, we use the weighted F1-score, a metric that computes the F1-score for each class and averages them according to their support (i.e., the number of true instances per class). Since the labels are highly imbalanced, we have used the weighted F1-score that considers both the correctly predicted positive labels $(+1)$ and negative labels $(0)$.  {To evaluate the stability of DP-LoRA training method, we plot training and validation loss curves across training epochs. As shown in Figure} \ref{fig:bertsmall_loss},  {the validation loss closely follows the training loss without divergence, indicating that the injected DP noise acts as a regularizer and prevents memorization. Unlike non-private fine-tuning, which shows early overfitting, DP-LoRA exhibits smoother convergence and improved generalization.}

\subsection{Effect of DP on classification performance} In this experimental section, we compare the effect of differentially private fine-tuning in different privacy regimes compared to the non-fine-tuned models and models fine-tuned without any privacy constraints. 
From Figure \ref{fig:results_table}, and Figure \ref{fig:f1score} we observe that the performance of ALBERT-base, BERT-small, and BERT-medium models across different training conditions, including off-the-shelf (non-fine-tuned), differential privacy (DP-SGD) with varying epsilon values, and non-private fine-tuning methods. When fine-tuned with DP-SGD, the models show improved performance for both the datasets, as epsilon increases, indicating a clear privacy-utility tradeoff. In contrast, non-private fine-tuning significantly boosts accuracy: on MIMIC-CXR, non-private LoRA achieves 0.901 for BERT models and 0.891 for ALBERT-base, while full fine-tuning results in the highest accuracy across all models, with BERT-medium reaching 0.968. On CT-RATE dataset, the non-private LoRA achieves ~ 0.79 F1-score whereas the non-private full fine-tuning achieves 0.97. Overall, full fine-tuning without privacy provides the best performance, while DP-SGD enables comparable F1-score with strong privacy guarantees. We observe a higher F1-score on the MIMIC-CXR dataset compared to CT-RATE, primarily due to two factors. First, the reference standard labels for MIMIC-CXR were generated using CheXbert, a BERT-based model fine-tuned on the MIMIC dataset, which aligns closely with the domain and structure of the test data. Second, CT reports tend to be longer and more complex than CXR reports, introducing additional linguistic and semantic challenges for automated label extraction.
\subsection{Effect of LoRA Rank on Utility} Intuitively, as we increase the LoRA ranks, the predictive performance of the models gets better, achieving significantly better F1-scores.
The results in Figure \ref{fig:lora} highlight the critical role of LoRA rank (r) in navigating the privacy-utility trade-off during LLM fine-tuning. Under a fixed total privacy budget, increasing the LoRA rank (r) leads to a proportional increase in the number of trainable parameters. As a result, more noise must be added per training step to satisfy differential privacy constraints, which in turn degrades the model’s predictive performance, most notably reflected in lower F1 scores. This demonstrates that the choice of LoRA rank is not merely a tuning parameter but a key factor in balancing model utility and privacy. Careful calibration of the rank is therefore essential to optimize performance under stringent privacy guarantees.

\begin{figure}[h]
    \centering
    \includegraphics[scale=0.25]{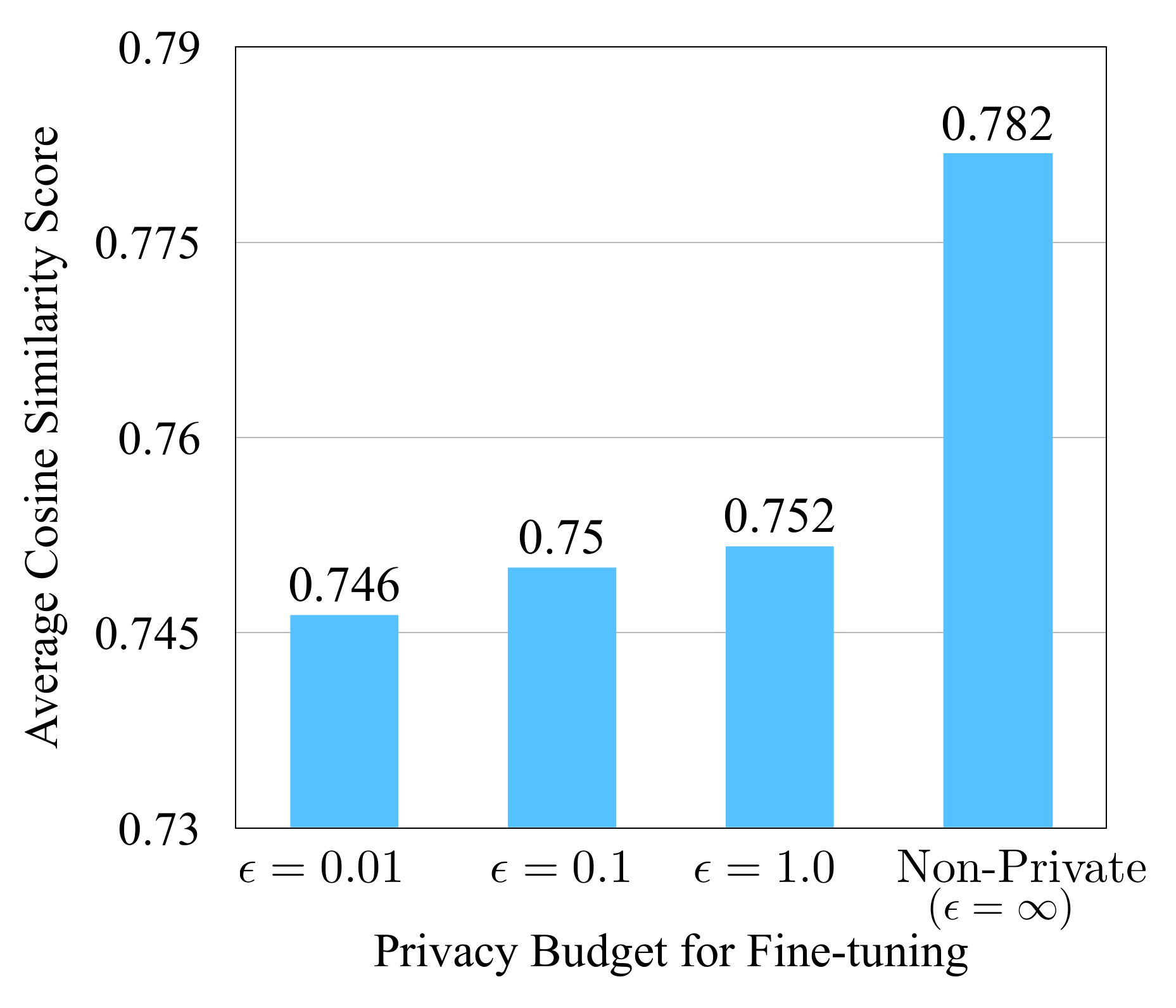}
    \caption{ {Average cosine similarity between ground-truth MIMIC-CXR reports and BERT-medium based fine-tuned model-generated completions under varying differential privacy budgets ($\epsilon$). Lower $\epsilon$ values (stronger privacy guarantees) result in reduced semantic similarity, indicating diminished memorization compared to the non-private fine-tuned model.}}
    \label{fig:avg_cosine}
    \vspace{-1em}
\end{figure}

\subsection{Impact of Differential Privacy on Semantic Memorization } {The empirical trend shown in Figure} \ref{fig:memorization_2}  {provides quantitative evidence that differentially private (DP) fine-tuning mitigates memorization in radiology report generation. Specifically, we measure the cosine-similarity between the ground-truth MIMIC-CXR report and the model-generated completion under varying privacy budgets. As the privacy budget decreases (i.e., stronger privacy guarantees with smaller $\epsilon$), the average cosine similarity systematically declines relative to the non-private baseline. The non-private fine-tuned model achieves the highest cosine similarity score, whereas progressively smaller $\epsilon$ values (e.g., $\epsilon = 1.0, 0.1, 0.01$) yield lower similarity scores, indicating reduced reconstruction fidelity to the original report. This trend is consistent with the theoretical objective of differential privacy, which constrains gradient updates and limits the model’s capacity to memorize training-specific content. Importantly, the reduction in similarity reflects controlled suppression of memorization patterns rather than arbitrary performance degradation, thereby providing an empirical privacy evaluation of the trained models. These results directly address concerns regarding the absence of privacy-specific metrics by demonstrating that DP fine-tuning measurably reduces semantic proximity to ground-truth reports while preserving clinically coherent generation, thus achieving improved privacy preservation.}

\section{Discussion} LLMs offer a scalable and efficient solution for parsing free-text radiology reports, enabling both structured data extraction and multi-abnormality labeling. These capabilities are essential for building high-quality labeled datasets and improving the efficiency of downstream diagnostic tasks, especially tasks that are otherwise difficult to scale through manual annotation. In this study, we propose a privacy-preserving fine-tuning framework that integrates Differential Privacy (DP) with Low-Rank Adaptation (LoRA) for multi-abnormality classification from free-text radiology reports. By combining parameter-efficient adaptation with formal privacy guarantees, our approach directly addresses the privacy–utility trade-off inherent in deploying LLMs within healthcare environments. Through extensive experimentation on the MIMIC-CXR \cite{johnson_mimic-cxr_2019} and CT-RATE  datasets, we demonstrate that the proposed DP-LoRA framework achieves competitive classification accuracy while maintaining rigorous differential privacy guarantees. The results clearly illustrate that although stricter privacy budgets reduce predictive performance, moderate privacy regimes yield weighted F1-scores comparable to non-private baselines. These findings confirm that privacy protection and model utility can coexist effectively when fine-tuning large language models on sensitive clinical text data.  {DP-LoRA is particularly suitable for low-resource environments because only low-rank adapter parameters are trained while the backbone model remains frozen. This reduces memory consumption and GPU requirements compared to full fine-tuning. In addition, differential privacy acts as an implicit regularizer, allowing smaller models to achieve competitive performance. Therefore, the framework can be deployed in hospital settings without large-scale compute infrastructure.}

Compared with prior work \cite{erberk_uslu_nlp-powered_2024}, which focused on fine-tuning LLMs for radiology report analysis without privacy considerations, our method achieves comparable weighted F1-scores while simultaneously providing formal DP guarantees. A key contribution of this study lies in systematically analyzing how privacy budgets $(\epsilon, \delta)$ and LoRA rank ($r$) jointly influence predictive performance. We observe a clear dependency between the number of trainable parameters and downstream utility. In non-private settings, increasing the LoRA rank consistently improves F1-scores due to enhanced fine-tuning capacity. However, under strict differential privacy constraints, increasing the number of trainable parameters amplifies gradient noise injection, thereby diminishing performance gains. This highlights the importance of carefully calibrating privacy budgets and adaptation rank to achieve an optimal privacy-utility tradeoff.

 {In practical deployments, the selection of the privacy budget $(\epsilon, \delta)$ should be guided by principled risk assessment rather than arbitrary thresholds. Prior work has emphasized that $\epsilon$ quantifies a bounded increase in individual privacy risk and must be interpreted within contextual, regulatory, and institutional constraints} \cite{hsu_differential_2014}.  {Empirical analyses of differentially private machine learning further demonstrate that the implications of $\epsilon$ depend on dataset sensitivity, population size, and adversarial assumptions} \cite{jayaraman2019evaluating}.  {Accordingly, privacy parameter selection in medical applications such as radiology report analysis should be grounded in broader risk-management frameworks rather than fixed numerical targets. Our evaluation across multiple privacy regimes provides empirical reference points that are consistent with established theoretical perspectives.}

Importantly, the notion of “strong performance” must be contextualized within clinical practice.  {In radiology report classification literature, weighted F1-scores between 0.80-0.95 are typically considered clinically usable depending on abnormality prevalence and decision-support role. In our setting, the model serves as a triage and pre-annotation tool rather than an autonomous diagnostic system. Therefore, F1-scores in the 0.85-0.90 range under privacy constraints indicate practical clinical utility while preserving patient confidentiality.}

Beyond quantitative performance, our findings underscore the practical viability of deploying lightweight models in real-world healthcare workflows. By differentially privately fine-tuning compact architectures such as BERT-small and ALBERT-base, we maintain competitive predictive performance while satisfying computational efficiency and confidentiality requirements critical for hospital environments. Although promising, this study has several limitations that motivate future research. First, although we demonstrate effectiveness on chest X-ray-based and CT-based free-text radiology reports, generalizability to broader clinical contexts and multi-modal datasets remains to be explored. Extending DP-LoRA to multi-modal learning scenarios that jointly model imaging and textual data represents an important direction. Second, although we motivate differential privacy through memorization risks, a deeper empirical evaluation of adversarial attack resilience and membership inference resistance would further strengthen deployment readiness. Third, our implementation relies on standard DP-SGD with Gaussian noise; more advanced strategies such as adaptive privacy budgeting, alternative noise mechanisms, or privacy amplification techniques may offer improved privacy–utility trade-offs.  {A promising future direction is combining DP-LoRA with federated learning to enable multi-center collaborative training without data sharing. Such integration would allow institutions to jointly train models while maintaining both local data confidentiality and formal differential privacy guarantees, making the framework suitable for large-scale clinical deployment.}

In summary, this work presents a differential privacy-preserving fine-tuning framework for multi-abnormality classification from radiology reports that is computationally efficient, empirically validated, and practically deployable. By integrating LoRA with differential privacy mechanisms, we demonstrate that strong privacy protection and competitive clinical performance can be achieved simultaneously. The framework is extensible to multi-organ, multi-abnormality classification, multi-modal medical data analysis, report generation, and summarization tasks. As future work, we plan to conduct a more comprehensive analysis of privacy vulnerabilities and explore robust privacy-preserving strategies tailored to large-scale, multi-institutional, multi-modal medical applications.

\section{Conclusions}  {This paper presents an adaptation of  differentially privacy for fine-tuning of LLMs for multi-abnormality classification from free-text radiology reports using DP-LoRA mechanism. Experiments on MIMIC-CXR and CT-RATE demonstrate that strong privacy guarantees can be achieved with only moderate reduction in classification accuracy compared to the non-private setting. The results confirm that privacy-preserving adaptation of language models is feasible for real clinical deployment. Future work will extend the framework to multi-modal medical data, federated and split learning based frameworks and advanced privacy mechanisms.}

\section{Acknowledgment}This work was supported by NIH Award R01-CA261457-01A1 and also by the US Department of Energy, Office of Science, Office of Advanced Scientific Computing under Award Number DE-SC-ERKJ422. US NSF under Grants CCF-2100013, CNS2209951, CNS-2317192.

\EOD

\end{document}